\documentclass[10pt,letterpaper,english,twocolumn, conference]{IEEEtran}
\usepackage[T1]{fontenc}
\usepackage[latin9]{luainputenc}
\usepackage{amsmath}
\usepackage{amsthm}
\usepackage{amssymb}
\usepackage{graphicx}
\usepackage{enumerate}

\makeatletter

\@ifundefined{pageheight}{\let\pageheight\pdfpageheight}{}
\@ifundefined{pagewidth}{\let\pagewidth\pdfpagewidth}{}
\pageheight\paperheight
\pagewidth\paperwidth

\theoremstyle{plain}

\theoremstyle{plain}

\theoremstyle{remark}

\@ifundefined{date}{}{\date{}}
\usepackage{subfigure}
\usepackage{epstopdf}
\usepackage{cite}
\usepackage{balance}

\makeatother

\usepackage{babel}
\providecommand{\lemmaname}{Lemma}
\providecommand{\remarkname}{Remark}
\providecommand{\theoremname}{Theorem}

\begin{document}

\title{Multi-hop Links Quality Analysis of 5G Enabled Vehicular Networks}


%
\author{\IEEEauthorblockN{Shikuan Li\IEEEauthorrefmark{1},
Zipeng Li\IEEEauthorrefmark{1},
Xiaohu Ge\IEEEauthorrefmark{1},
Jing Zhang\IEEEauthorrefmark{1},
Minho Jo\IEEEauthorrefmark{2}
}
\IEEEauthorblockA{\IEEEauthorrefmark{1}School of Electronic Information and Communications, Huazhong University of Science and Technology, China}
\IEEEauthorblockA{\IEEEauthorrefmark{2}Department of Computer Convergence Software, Korea University, South Korea}
\IEEEauthorblockA{Corresponding author: Xiaohu Ge. Email:
\IEEEauthorrefmark{1}\{xhge\}@hust.edu.cn}}


\maketitle
\begin{abstract}
With the emerging of the fifth generation (5G) mobile communication systems, millimeter wave transmissions are believed to be a promising solution for vehicular networks, especially in vehicle to vehicle (V2V) communications. In millimeter wave V2V communications, different vehicular networking services have different quality requirements for V2V multi-hop links. To evaluate the quality of different V2V wireless links, a new link quality indicator is proposed in this paper considering requirements of the real-time and the reliability in V2V multi-hop links. Moreover, different weight factors are configured to reflect the different requirements of different types of services on real-time and reliability in the new quality indicator. Based on the proposed link quality indicator, the relationship between V2V link quality and one-hop communication distance under different vehicle densities is analyzed in this paper. Simulation results indicate that the link quality is improved with the increasing of vehicle density and there exists an optimal one-hop communication distance for the link quality when the vehicle density is fixed.
\end{abstract}

\section{Introduction}

\label{sec1}

In recent years, the vehicular network is considered to be one of the most promising technologies for intelligent transportation system (ITS) \cite{Sun1,Karagiannis2,Ge3}. A secure and efficient Vehicle-to-Vehicle (V2V) communications is the key to support road safety and traffic efficiency applications in ITS. At the same time, with the emerging of fifth generation (5G) mobile communication system \cite{Ghosh3,Mastrosimone4,Ge5}, millimeter wave communications have been considered to be an effective technology in V2V communications. Millimeter wave links are adopted for wireless multi-hop communications among vehicles. Therefore, different information can be transmitted and shared among vehicles, which enabled all kinds of services and applications in vehicular networks \cite{Araniti5,Sichitiu6}. These services and applications can mainly be classified into two types. The first is the safety services which require an extremely low latency, such as the vehicle collision warning, obstacle detection and driving assistances. The other is the infotainment services which prefer link reliability to the latency, such as file transmissions, social entertainments and online services.

In V2V communications, how to select the best communication link to meet different requirements of all kinds of services has become a hot topic recently. Many schemes have been proposed to reduce transmission delay in V2V communications \cite{Hu7,Choi8,Stamatiou9,Li10}. In \cite{Hu7}, the multi-hop transmission delay was studied under different probability distributions of network nodes. Mathematical expressions of the average communication delay among vehicles were derived for interference-limited cases and noise-limited cases. Delay-optimal routing algorithms in V2V communications were investigated to select the communication link with the lowest transmission delay \cite{Choi8,Stamatiou9}. Safety-related messages broadcasting delay in V2V communications was analyzed in \cite{Li10}. The analytic result indicated that broadcasting delays were different when one-hop transmission ranges among vehicles were different, so vehicles can broadcast the safety-related messages more efficiently by selecting an optimal transmission range. But this result was only available in messages broadcasting scenarios and only the broadcasting delay was analyzed. However, some vehicular services focused more on the V2V link reliability than the time delay and can tolerate a large transmission delay. How to improve the connectivity probability has been investigated in several works \cite{Zhang11,Li12,Zhao13,Shao14,Sou15}. An analytical model supporting multi-hop relay based on V2V communications was proposed to analyze the relationship between connectivity probabilities and hops under different scenarios \cite{Zhang11}. In \cite{Li12}, an analytical model was proposed to analyze the successful transmission probabilities of safety messages in linear multi-hop V2V links. The connectivity probabilities of different communication modes, i.e. the unit disk model and the log-normal shadowing model were compared in \cite{Zhao13}. An analytical mode taking account of the vehicle mobility and the path loss of wireless channel was proposed to simulate real scenarios of vehicular networks. The connectivity characteristic of platoon-based V2V communication scenarios was studied in \cite{Shao14}, \cite{Sou15}. Analysis results showed that the connectivity probability can be significantly improved when there are platoons in the vehicular network.

However, most of these studies only considered the delay-optimal V2V links or the V2V links with the best connectivity probability. Very few study was done to optimize the transmission delay and multi-hop connectivity in one analytical model of vehicle networks. By taking both the transmission delay and the connectivity into consideration, in this paper we proposed a new link quality indicator to evaluate the V2V multi-hop link quality. Moreover, the relationships between the link quality and network parameters are investigated. Based on the proposed link quality indicator, optimal V2V multi-hop links could be realized in 5G enabled vehicular networks. The contributions and novelties of this paper are summarized as follows.

\begin{itemize}
\item Considering the multi-hop V2V transmission delay and connectivity, a new link quality indicator was proposed to evaluate the V2V multi-hop link quality. Different kinds of vehicular service requirements are reflected by modifying the weight factors of the transmission delay and connectivity in the link quality indicator.
\item The multi-hop transmission delay and the connectivity probability of millimeter wave based V2V communication are analyzed. The analytical expression of the V2V link quality indicator is derived.
\item Simulation results indicate that V2V multi-hop link strategies depend on V2V transmission delay, connectivity probability and weight factors. These results provide some useful guidelines for selection strategies of V2V communication links in vehicular networks.
\end{itemize}

The remainder of this paper is organized as follows. Section II describes the system model and V2V link quality evaluation model. The multi-hop V2V transmission delay and connectivity probability with different communication strategies are investigated in Section III. In Section IV, simulations results show the impact of V2V transmission delay, connectivity probability and weight factors on the V2V link quality. Finally, Section V concludes this paper.



\section{System Model}
\label{sec2}

\subsection{System scenarios}
\label{sec2-1}

As shown in Fig. 1, we consider the scenario of a single lane in urban area, where vehicles on the single lane can communicate with each other by millimeter wave communications. Assuming that there is a vehicle which wants to send some messages to another vehicle on the single lane. These messages include the driving statuses of vehicles, safety-related messages, voice messages, etc. Without loss of generality, we assume that the vehicle ${V_a}$ have some messages to be sent to the vehicle ${V_b}$ where is $L$ meters away from ${V_a}$. In most cases, since the millimeter wave communication range is generally less than 200 meters [3], ${V_a}$ can not communicate with ${V_b}$ directly. Hence a multi-hop V2V link need to be performed to transmit these messages when $L$ is larger than 200 meters.

\begin{figure*}
  \centering
  \includegraphics[width=17cm,draft=false]{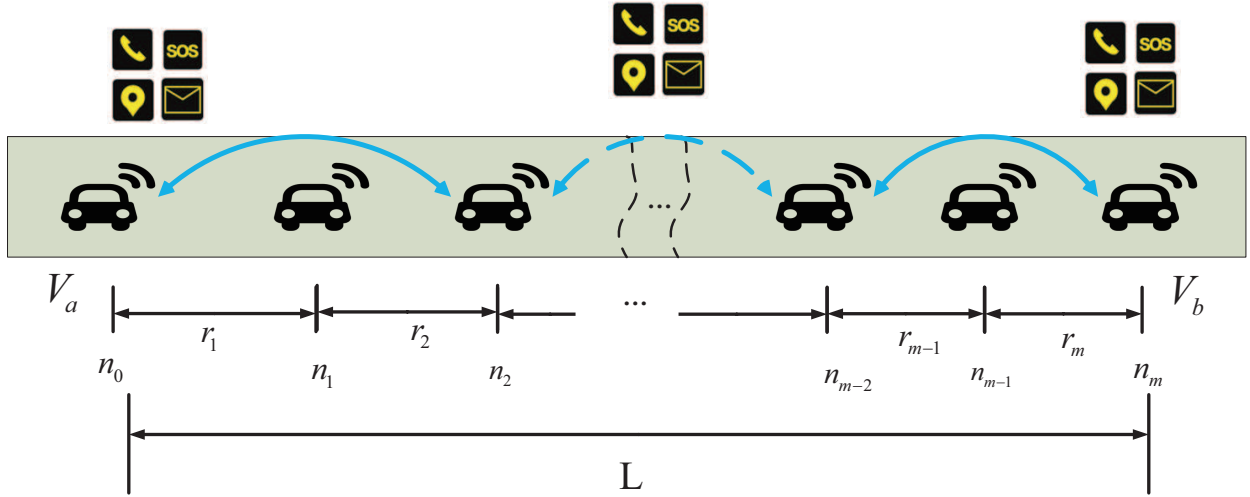}\\
  \caption{\small System model.}
  \label{fig1}
\end{figure*}

According to the traffic flow model\cite{Zhang11}, the probability that there are   vehicles on the   meter road is expressed as

\begin{equation}
P(n,L) = \frac{{{{(\rho L)}^n}{e^{ - \rho L}}}}{{n!}},
\end{equation}

where $\rho $ is the traffic density, defined as the number of cars per meter. Assuming that $m$ is the total number of cars between ${V_a}$ and ${V_b}$ and $m$ is subject to Poisson distribution with density of $\rho $. Without loss of generality, ${V_a}$ is denoted as node ${n_0}$ and the ${i^{th}}$ vehicle away from vehicle ${V_a}$ is denoted as the node ${n_i}$. Obviously, vehicle ${V_b}$ is denoted as node ${n_m}$ and the distance between node ${n_i}$ and ${n_0}$ is denoted as ${x_i}$ and ${x_m} = L$.

\subsection{V2V multi-hop link quality indicator}
\label{sec2-2}

To ensure requirements of messages satisfying the transmission delay and the transmission success probability, a multi-hop V2V link with the best link quality should be selected. To evaluate the link quality, the multi-hop link quality indicator ${Q^ * }$ is defined as

\begin{equation}
{Q^ * } = \alpha {\rm P} + \beta D,
\end{equation}

where $P \in \left[ {0,1} \right]$ represents the connect probability of ${V_a}$ and ${V_b}$, the weight factor $\alpha $ is used to indicate requirement extent of the connect probability, the weight factor $\beta $ is used to indicate requirement extent of transmission delay. Weight factors $\alpha $ and $\beta $ should meet the formula requirement as $\left\{ {\alpha  + \beta  = 1\left| {\alpha ,\beta  \in \left[ {0,1} \right]} \right.} \right\}$. Symbol $D$ is the indicator of transmission delay, the definition of $D$ is given by

\begin{equation}
D = 1 - \frac{{T\left( \phi  \right)}}{{{T_{\max }}}},
\end{equation}

where $T\left( \phi  \right)$ is the message transmission delay on the multi-hop link $\phi $, ${T_{\max }}$ is the max transmission delay that can be tolerated. The value range of $D$ in equation (3) is $D \in \left( { - \infty ,1} \right)$. When $D$ is a negative number it means that the message transmission delay on the multi-hop link $\phi $ is larger than ${T_{\max }}$. If vehicle ${V_a}$ send the message to ${V_b}$ through a multi-hop link $\phi $, the link quality of multi-hop link $\phi $ is given as ${Q^ * }\left( \phi  \right) \in \left( { - \infty ,1} \right]$. Substituting equation (3) into equation (2), ${Q^ * }\left( \phi  \right)$ is derived by

\begin{equation}
{Q^ * }\left( \phi  \right) = \alpha P + \beta \left( {1 - \frac{{T\left( \phi  \right)}}{{{T_{\max }}}}} \right).
\end{equation}

As for different services in vehicular networks, different transmission delay and connectivity requirements are needed. Therefore, selecting a multi-hop link with optimal link quality to satisfy these requirements will make the vehicular network more efficient and reliable. For a fixed $\alpha $ and $\beta $, the multi-hop link with a larger link quality ${Q^ * }$ means that the performance of this multi-hop link is more suitable for a specific type of message dissemination. In Section III, transmission delay and link connectivity probability will be evaluated in details, impacts of network parameters on link quality will be investigated to help selecting the optimal V2V link under different vehicular network scenarios.

\section{Transmission delay and connectivity probability of V2V links}
\label{sec3}

\subsection{Transmission delay of V2V links}
\label{sec3-1}

To satisfy different requirements of vehicular network services on messages transmission delay and transmission success probability, relay vehicles should be selected accurately to optimizing the link quality in V2V multi-hop links. Without loss of generality, vehicle nodes $\left\{ {{c_1},{c_2}, \cdots ,{c_{k - 1}}} \right\}$ are denoted as relay nodes which are selected from $\left\{ {{n_1},{n_2}, \cdots ,{n_{m - 1}}} \right\}$ to transmit messages. The transmission delay of V2V multi-hop link $\phi \left\{ {{c_0},{c_1}, \cdots ,{c_k}} \right\}$ is expressed as

\begin{equation}
{T_\phi } = \left( {k - 1} \right){T_{pro}} + \sum\limits_{i = 1}^k {{T_{hop}}_j},
\end{equation}

where ${T_{hop}}_j$ represents the transmission delay of messages between relay node ${c_{j - 1}}$ and ${c_j}$, ${T_{pro}}$ represents the processing time by relay node ${c_j}$.

\begin{enumerate}[a)]
\item One hop transmission delay  in V2V links
\end{enumerate}

Due to the characteristic of millimeter wave, the path loss $PL$ in any hop of $\phi $ should not be ignored and is given as \cite{Ghosh3}

\begin{equation}
PL[dB]({r_j}) = 69.6 + 20.9\log ({r_j}) + \xi ,\xi  \sim N\left  (  {0,{\sigma ^2}} \right),
\end{equation}

where ${r_j}$ is denoted as the distance between vehicle node ${c_{j - 1}}$ and ${c_j}$, $\xi $ is the shadow fading coefficient , and $\sigma $ is the standard deviation of shadow fading, which is estimated as 5 dB. The transmit power plus antenna gains of vehicle nodes is denoted as ${P_{tx}}$ and ${N_0}$ represents the Gaussian white noise power density. The single hop transmission is regarded as success if the received signal-to-noise ratio (SNR) of relay vehicle is larger than a given threshold $\theta $. Hence, the transmission success probability in one hop is defined as

\begin{equation}
{P_{hop}}_j = {\rm P}\left( {PL \le {P_{tx}}(dB) - \theta (dB) - {N_0}{W_{mmWave}}(dB)} \right),
\end{equation}

where ${W_{mmWave}}$ is the millimeter wave bandwidth. Substituting (6) into (7), the success probability of one hop transmission in a single slot is derived by equation(8), where $erf()$ is the error function and \[\eta \left( {{r_j}} \right) = {P_{tx}} - \theta  - {N_0}{W_{mmWave}} - 69.6 - 20.9{\log _{10}}{r_j}.\]

\begin{figure*}[!t]
\begin{equation}
\begin{gathered}
\begin{array}{l}
{P_{hopj}} = {\rm P}(\xi  \le {P_{tx}} - \theta  - {N_0}{W_{mmWave}} - 69.6 - 20.9{\log _{10}}{r_j})\\
\\
{\kern 1pt} {\kern 1pt} {\kern 1pt} {\kern 1pt} {\kern 1pt} {\kern 1pt} {\kern 1pt} {\kern 1pt} {\kern 1pt} {\kern 1pt} {\kern 1pt} {\kern 1pt} {\kern 1pt} {\kern 1pt} {\kern 1pt} {\kern 1pt} {\kern 1pt} {\kern 1pt} {\kern 1pt} {\kern 1pt} {\kern 1pt} {\kern 1pt} {\kern 1pt}  = \frac{1}{2}\left( {1 + erf\left( {\frac{{\eta \left( {{r_j}} \right)}}{{\sqrt 2 \sigma }}} \right)} \right)
\end{array}
\end{gathered}
\end{equation}
\end{figure*}

Without loss of generality, let ${t_{slot}}$ be the time cost of a single slot, due to the large bandwidth of millimeter wave, we can assume that for a successful transmission, the messages can be sent within a single time slot, so the one hop transmission delay ${T_{hop}}_j$ can be derived by using the time cost of a time slot over the success transmission probability expressed as

\begin{equation}
\begin{array}{l}
{T_{hopj}} = \frac{{{t_{slot}}}}{{{P_{hopj}}}}\\
\\
{\kern 1pt} {\kern 1pt} {\kern 1pt} {\kern 1pt} {\kern 1pt} {\kern 1pt} {\kern 1pt} {\kern 1pt} {\kern 1pt} {\kern 1pt} {\kern 1pt} {\kern 1pt} {\kern 1pt} {\kern 1pt} {\kern 1pt} {\kern 1pt} {\kern 1pt} {\kern 1pt} {\kern 1pt} {\kern 1pt} {\kern 1pt} {\kern 1pt}  = \frac{{2{t_{slot}}}}{{1 + erf\left( {\frac{{\eta \left( {{r_j}} \right)}}{{\sqrt 2 \sigma }}} \right)}}
\end{array},
\end{equation}

\begin{enumerate}[b)]
\item Connection strategy in V2V links
\end{enumerate}

The transmission delay on multi-hop link $\phi $ shows that the total hop number $k$ is a key network parameter which determines the transmission delay. As for a larger hop number, the distance between two adjacent relay vehicles is supposed to be shorter, which means that an optimal link connection strategy should be selected to determine the one hop communication distance. For any relay vehicle ${c_j}$ whose distance from ${n_0}$ is ${x_j}$ in the V2V multi-hop links, the connection strategy which determines to select the next relay vehicle ${c_{j + 1}}$ is configured by

\begin{equation}
{c_{j + 1}} = \mathop {argmin}\limits_{{n_v} \in \left\{ {{n_0}, \cdots ,{n_m}} \right\}} \left| {{x_v} - \left( {{x_j} + r} \right)} \right|,
\end{equation}

where $r$ is the average one hop communication distance, ${x_v}$ is the distance between ${n_v}$ and vehicle ${n_0}$. The connection strategy is that for a relay vehicle ${c_j}$, the vehicle node ${n_v}$ is always selected as the next relay vehicle ${c_{j + 1}}$ based on the result of (10).

\begin{enumerate}[c)]
\item Multi-hop link time delay in V2V links
\end{enumerate}

When all relay vehicles follow the connection strategy in (10), the transmission delay ${T_\phi }$ in (5) is simplified as

\begin{equation}
T\left( \phi  \right) = E\left( {k{T_{hopj}}} \right) + \left( {E\left( k \right) - 1} \right){T_{pro}},
\end{equation}

where $E\left( k \right) = \left\lceil {L/r} \right\rceil $ represents the expectation of the hop number in the multi-hop link $\phi $ when the average one hop communication distance is $r$. $\left\lceil {L/r} \right\rceil $ represents the minimum integer that greater than or equal to $L/r$. Substituting (8) into (11), the total transmission delay $T\left( \phi  \right)$ on the V2V multi-hop $\phi $ is derived by

\begin{equation}
T\left( \phi  \right) = \frac{{2\left\lceil {{\raise0.5ex\hbox{$\scriptstyle L$}
\kern-0.1em/\kern-0.15em
\lower0.25ex\hbox{$\scriptstyle r$}}} \right\rceil {t_{slot}}}}{{\left( {1 + {\rm M}\left( r \right)} \right)}} + \left\lceil {{\raise0.5ex\hbox{$\scriptstyle L$}
\kern-0.1em/\kern-0.15em
\lower0.25ex\hbox{$\scriptstyle r$}} - 1} \right\rceil {T_{pro}},
\end{equation}

where ${\rm M}\left( r \right) = erf\left( {\frac{{\eta \left( r \right)}}{{\sqrt 2 \sigma }}} \right)$ and $\eta \left( r \right) = {P_{tx}} - \theta  - {N_0}{W_{mmWave}} - 69.6 - 20.9{\log _{10}}r$. Based on (12), the transmission delay $T\left( \phi  \right)$ is determined by the distance $L$ between vehicle ${V_a}$ and ${V_b}$, one hop average communication distance $r$, the transmit power ${P_{tx}}$ and the SNR threshold $\theta $.

\subsection{V2V transmission connectivity probability}
\label{sec3-2}

In this paper, the unit disk communication model is introduced to determine the connection probability between the relay vehicle node ${c_j}$ and ${c_{j + 1}}$. For the unit disk model, any vehicle pair is able to be connected through millimeter wave communications if the distance between each pair is less than the coverage radius. Thus, the connectivity probability between vehicle node ${c_j}$ and ${c_{j + 1}}$ is expressed as

\begin{equation}
{P_c} = \left\{ {\begin{array}{*{20}{c}}
{1,}&{{r_v} \le {R_v}}\\
{0,}&{otherwise}
\end{array}} \right.,
\end{equation}

where ${r_v}$ represents the distance of vehicle node ${c_j}$ and ${c_{j + 1}}$, the coverage radius of millimeter wave is fixed as ${R_v}$.

Since the number of vehicles is subject to Poisson distribution with the parameter $\rho $, the distances ${r_i}\left( {i = 1,2, \cdots ,m} \right)$ between adjacent vehicle nodes are subject to an exponential distribution with the parameter $\rho $. Hence, the cumulative distribution function (cdf) of ${r_i}$ is given by

\begin{equation}
P\left\{ {{r_i} \le {r_0}} \right\} = 1 - {e^{ - \rho {r_0}}},{r_0} \ge 0.
\end{equation}

Due to the memoryless property of Poisson distribution, the vehicle number which between two connected vehicle nodes in link $\phi $ is also subject to the Poisson distribution with the same parameter $\rho $. Without loss of generality, let $s$ be the vehicle number within one hop communication distance ${r_v}$, the expectation of vehicle number can be expressed as $E\left( s \right) = \rho {r_v}$. Obviously, ${r_v}$ is the sum of $s$ independent exponential variables with the same parameter $\rho $, which means that ${r_v}$ is subject to the Erlang distribution with shape parameter $s$. Therefore, the cumulative distribution function (cdf) of ${r_v}$ is given by

\begin{equation}
P\left\{ {{r_v} \le {r_0}} \right\} = F\left( {{r_0}} \right) = 1 - {e^{ - \rho {r_0}}}\sum\limits_{i = 0}^{s - 1} {\frac{{{{\left( {\rho {r_0}} \right)}^i}}}{{i!}}} ,{r_0} \ge 0.
\end{equation}

According to (13) and (15), the connectivity probability $P$ of V2V multi-hop link $\phi \left\{ {{c_0},{c_1}, \cdots ,{c_k}} \right\}$ between vehicle ${V_a}$ and ${V_b}$ is derived by (16), where ${r_v}\left( {v = 1,2, \cdots ,k} \right)$ is the one hop communication distance, $r = E\left( {{r_v}} \right)$ is the average one hop communication distance.

\begin{figure*}[!t]
\begin{equation}
\begin{gathered}
\begin{array}{l}
P = {\mathop{\rm P}\nolimits} \left( {{r_1} \le {R_v}} \right){\mathop{\rm P}\nolimits} \left( {{r_2} \le {R_v}} \right) \cdots {\mathop{\rm P}\nolimits} \left( {{r_k} \le {R_v}} \right)\\
\\
{\kern 1pt} {\kern 1pt} {\kern 1pt} {\kern 1pt} {\kern 1pt} {\kern 1pt} {\kern 1pt} {\kern 1pt} {\kern 1pt} {\kern 1pt} {\kern 1pt}  = \prod\limits_{i = 1}^{{\rm E}\left( k \right)} {{\mathop{\rm P}\nolimits} \left( {E\left( {{r_v}} \right) \le {R_v}} \right)} \\
\\
{\kern 1pt} {\kern 1pt} {\kern 1pt} {\kern 1pt} {\kern 1pt} {\kern 1pt} {\kern 1pt} {\kern 1pt} {\kern 1pt}  = {\left( {1 - {e^{ - \rho {R_v}}}\sum\limits_{i = 0}^{\left\lceil {\rho r} \right\rceil  - 1} {\frac{{{{\left( {\rho {R_v}} \right)}^i}}}{{i!}}} } \right)^{\left\lceil {L/r} \right\rceil }}
\end{array}
\end{gathered}
\end{equation}
\end{figure*}

Since the expressions of V2V multi-hop transmission delay $T\left( \phi  \right)$ and connectivity probability $P$ have been derived, substituting (12) and (16) into (4), the V2V multi-hop link quality ${Q^ * }\left( \phi  \right)$ is derived by (17).

\begin{figure*}[!t]
\begin{equation}
\displaystyle
\begin{array}{l}
{Q^ * }\left( \phi  \right) = \alpha P + \beta \left( {1 - \frac{{T\left( \phi  \right)}}{{{T_{\max }}}}} \right)\\
\\
 = \alpha {\left( {1 - {e^{ - \rho {R_v}}}\sum\limits_{i = 0}^{\left\lceil {\rho r} \right\rceil  - 1} {\frac{{{{\left( {\rho {R_v}} \right)}^i}}}{{i!}}} } \right)^{\left\lceil {L/r} \right\rceil }} + \beta \left( {1 - \frac{{\frac{{2\left\lceil {{\raise0.5ex\hbox{$\scriptstyle L$}
\kern-0.1em/\kern-0.15em
\lower0.25ex\hbox{$\scriptstyle r$}}} \right\rceil {t_{slot}}}}{{\left( {1 + {\rm M}\left( r \right)} \right)}} + \left\lceil {{\raise0.5ex\hbox{$\scriptstyle L$}
\kern-0.1em/\kern-0.15em
\lower0.25ex\hbox{$\scriptstyle r$}} - 1} \right\rceil {T_{pro}}}}{{{T_{\max }}}}} \right)
\end{array}
\end{equation}
\end{figure*}

\section{Numerical Results and Discussion}
\label{sec4}

In this section, the V2V multi-hop link quality with different vehicles densities and different average one hop communication distances are numerically analyzed. The different requirements of different vehicular network services on messages transmission delay and transmission success probability are reflected by weight factors. Some default parameters are configured as follows:
$\rho  = 0.05 \sim 0.25$, $L = 1000m$, ${R_v} = 100m$, ${t_{slot}} = 50\mu s$, ${T_{pro}} = 20\mu s$, ${T_{\max }} = 20ms$, ${P_{tx}} = 30dBm$, ${W_{mmWave}} = 200MHz$, ${N_0} =  - 174dBm/Hz$ \cite{Ghosh3,Zhang11}.

\begin{figure}
  \centering
  \includegraphics[width=8.5cm,draft=false]{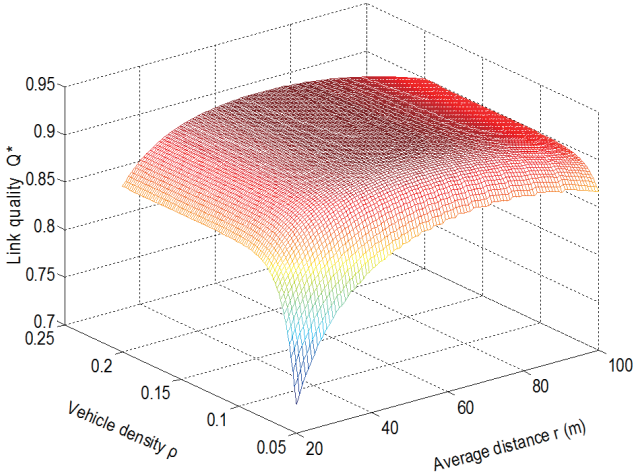}\\
  \caption{\small V2V multi-hop link quality with respect to vehicles density $\rho $ and average one hop communication distance $r$ ( $\alpha  = \beta  = 0.5$ ).}
  \label{fig2}
\end{figure}

Fig. 2 plots the V2V multi-hop link quality as a function of vehicles density $\rho $ and average one hop communication distance $r$ for the situation of fixed weight factors $\alpha  = \beta  = 0.5$. As shown in Fig. 2, multi-hop link quality ${Q^*}$ is a monotonically increasing function of vehicle density $\rho $. When the vehicle density is fixed, with the increase of average communication distance, multi-hop link quality ${Q^*}$ first increases and then deceases, which implies there exists an optimal average communication distance conducing to the best link quality.

\begin{figure}
  \centering
  \includegraphics[width=8.5cm,draft=false]{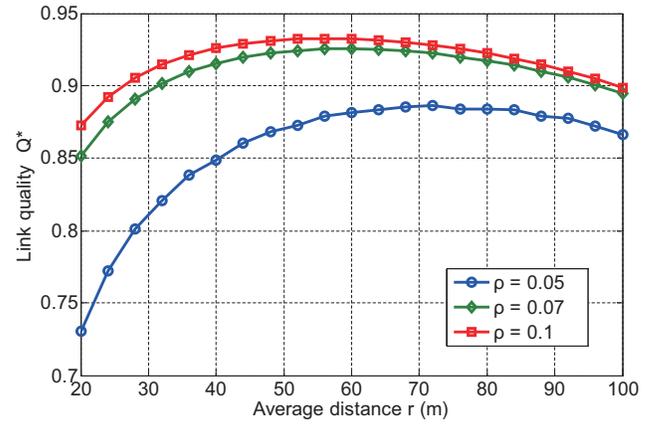}\\
  \caption{\small V2V multi-hop link quality with respect to different average one hop communication distance $r$ ( $\alpha  = \beta  = 0.5$ ).}
  \label{fig3}
\end{figure}

Fig. 3 shows V2V multi-hop link quality with respect to different average one hop communication distance $r$. Without loss of generality, the weight factors $\alpha $ and $\beta $ are both configured as $\alpha  = \beta  = 0.5$ in Fig. 3. it is shown that there exists an optimal average one hop communication distance to achieve the best link quality. The reason is that when the vehicle density is fixed, with a relatively small value of $r$, the relay vehicles will communicate with a vehicle that has a relatively close distance. Hence, this V2V multi-hop link will have a large of hop numbers. In this case, a relatively small value of $r$ conduces to a large connectivity probability. But the transmission delay increases with the increase of the hop number. Hence, the quality of this link is relatively low. On the other hand, a multi-hop link with a relatively large value of $r$ leads to a low transmission delay but the connectivity probability is relatively low. Hence, there exists a link with an optimum average distance which leads to the best link quality by trading off the transmission delay and connectivity probability. It is worth noting that with a different vehicle density, the optimum average distance is different. This result is helpful to find the best V2V communication strategy in vehicular networks.

\begin{figure}
  \centering
  \includegraphics[width=8.5cm,draft=false]{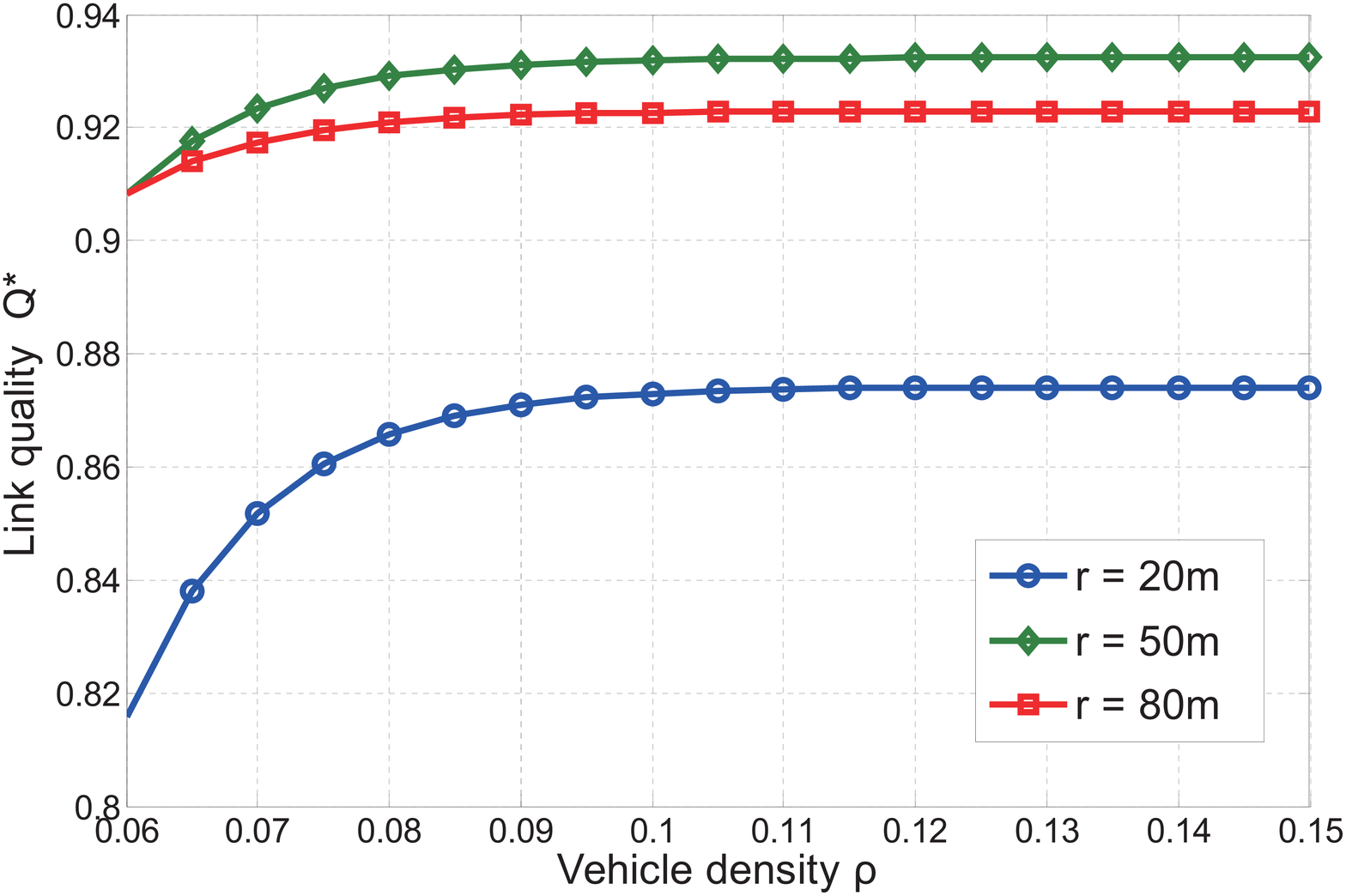}\\
  \caption{\small V2V multi-hop link quality with respect to different vehicle density $\rho $ ( $\alpha  = \beta  = 0.5$ ).}
  \label{fig4}
\end{figure}

In Fig. 4, the effect of vehicles density $\rho $ on V2V multi-hop link quality ${Q^*}$ is investigated. The weight factors are both fixed at $\alpha  = \beta  = 0.5$. As shown in Fig. 4, when one hop communication distance $r$ is fixed, the link quality increases with the increase of the vehicle density. The reason is that a large vehicle density implies there are more optional vehicles to be relay vehicle nodes. Hence, the connectivity probability increases with the increase of vehicle density.

 \begin{figure}
  \centering
  \includegraphics[width=8.5cm,draft=false]{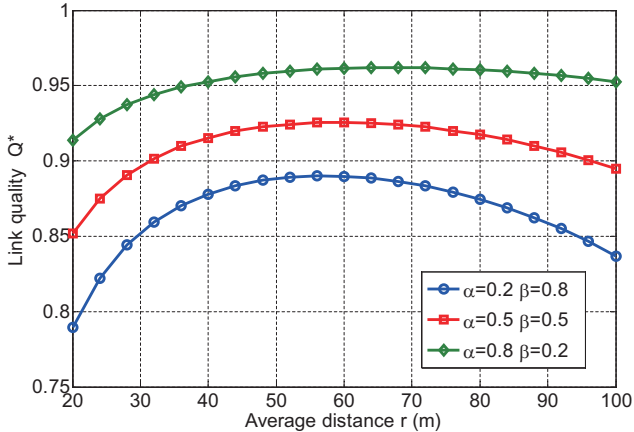}\\
  \caption{\small  V2V multi-hop link quality with respect to different weight factors $\alpha $ and $\beta $ ( $\rho  = 0.07$ ).}
  \label{fig5}
\end{figure}

Fig. 5 shows V2V multi-hop link quality with respect to different weight factors $\alpha $ and $\beta $. Without loss of generality, the vehicle density is configured as $\rho  = 0.07$ in Fig. 5, V2V multi-hop links have different link qualities when weight factors of transmission delay and transmission success probability are different. These results indicate that with different requirements of different vehicular network services on messages transmission delay and transmission success probability, the multi-hop link quality and the optimal average communication distance are different.

\section{Conclusion}
\label{sec5}

In this paper, a new link quality indicator has been proposed to evaluate multi-hop link quality by taking the transmission delay and the connectivity both into consideration. Furthermore, the analytical expressions of V2V multi-hop transmission delay and connectivity probability have been derived. Based on these, relationships between the link quality and some network parameters have been analyzed and numerical results shows that V2V link quality increases with the increase of the vehicle density, and there exists a optimal communication distance corresponding to the requirement of vehicular network services. These results indicated that when network parameters changes, the V2V multi-hop link quality can be improved by adjusting the link strategies. These results can also provide some useful guidelines for selection strategies of V2V communication links in vehicular networks.

\section*{Acknowledgment}

 The author would like to acknowledge the support from the NSFC Major International Joint Research Project under the grant 61210002, the Hubei Provincial Science and Technology Department under the grant 2016AHB006, the Fundamental Research Funds for the Central Universities under the grant 2016AHB006. This research is partially supported by EU FP7-PEOPLE-IRSES, project acronym CROWN (grant no. 610524).





%

\end{document}